\documentclass[twocolumn,superscriptaddress,showpacs,preprintnumbers,nofootinbib,amsmath,amssymb,floatfix,aps]{revtex4-1}
\usepackage{epsfig}
\usepackage{graphicx}
\usepackage{bm}
\usepackage{colordvi}
\usepackage[pagebackref]{hyperref}
\hypersetup{
    colorlinks=true, 
    linktoc=all,     
    linkcolor=blue,  
    citecolor=blue
}

\def\lsim{\buildrel < \over {_{\sim}}}

\newcommand{\beq}{\begin{equation}}
\newcommand{\eeq}{\end{equation}}
\newcommand{\be}{\begin{eqnarray}}
\newcommand{\ee}{\end{eqnarray}}

\begin{document}
\title{Transport properties of the Fermi hard-sphere system}
\author{Angela Mecca}
\affiliation{Dipartimento di Fisica,``Sapienza'' Universit\`a di Roma, I-00185 Roma, Italy}
\affiliation{INFN, Sezione di Roma, I-00185 Roma, Italy} 
\author{Alessandro Lovato}
\affiliation{Argonne Leadership Computing Facility, Argonne National Laboratory, Argonne, IL 60439, USA}
\affiliation{Physics Division, Argonne National Laboratory, Argonne, IL-60439, USA}
\author{Omar Benhar}
\affiliation{INFN, Sezione di Roma, I-00185 Roma, Italy}
\affiliation{Dipartimento di Fisica,``Sapienza'' Universit\`a di Roma, I-00185 Roma, Italy}
\author{Artur Polls}
\affiliation{Departament d'Estructura i Constituents de la Mat\`eria and Institut de Ci\`encies del Cosmos. E-08028 Barcelona, Spain}
\date{\today}
\begin{abstract}
The transport properties of neutron star matter play an important role in many astrophysical processes. We report the results
of a calculation of the shear viscosity and thermal conductivity coefficients of the hard-sphere fermion system of degeneracy $\nu$=2, 
that can be regarded as a model of pure neutron matter. Our approach is based on the effective interaction obtained from the formalism of 
correlated basis functions and the cluster expansion technique. The resulting transport coefficients show a strong sensitivity to the 
quasiparticle effective mass, reflecting the effect of second order contributions to the self-energy that are not taken into account in nuclear matter studies
available in the literature. 
 \end{abstract}
\pacs{24.10.Cn, 21.30.Fe}
\maketitle

\section{Introduction}
\label{intro}

The transport properties of nuclear matter are known to be relevant to a variety of astrophysical processes, such as neutron star cooling \cite{cooling} 
and the gravitational-wave-driven Chandrasekhar-Friedman-Schutz instability of rapidly rotating stars \cite{AndKok}. 

Following the seminal papers of Flowers and Itoh \cite{transpI,transpII}, several calculations of the thermal conductivity and 
shear viscosity coefficients of nuclear matter have been carried out using the formalism developed by Abrikosov and 
Khalatnikov \cite{ak} within the conceptual framework of  Landau's theory of normal Fermi liquids \cite{baym-pethick}.
In these studies, the nucleon effective masses and the in-medium nucleon-nucleon scattering cross sections\textemdash needed for the determination 
of the transport coefficients within the approach of Ref.~\cite{ak}\textemdash have been obtained using effective nucleon-nucleon interactions derived
from either G-matrix perturbation theory \cite{Gof4,lombardo,SBH} or the approach based on Correlated Basis Functions (CBF) and the
cluster expansion technique \cite{BV,Gof4}. 

The authors of Refs. \cite{Gof4,lombardo,SBH,BV} computed the effective masses within the Hartree-Fock 
approximation, corresponding to first order of the perturbative expansion in powers of matrix elements of the effective 
interaction\footnote{The scheme based on G-matrix and the Hartee-Fock approximation is generally referred to as 
Bruenker-Hartree-Fock approximation.}.
In Ref. \cite{paper1}, hereafter referred to as I, we have employed the CBF effective interaction approach to analyze the accuracy 
of this approximation for the fermion hard-sphere system of degeneracy $\nu$=4, corresponding to spin and isospin degrees of freedom.
The results reported in I show that the inclusion of the energy-dependent second order contributions to the proper self-energy, while leading to small 
modifications of the quasiparticle spectrum, strongly affects both magnitude and density dependence of the effective mass. 
In the Hartee-Fock approximation, the ratio between the effective mass of a quasiparticle on the Fermi surface and the bare mass, $m^\star(k_F)/m$, 
is consistently less than unity and decreases monotonically with  density. The second order result, on the contrary,  turns out to be larger than one 
and monotonically increasing. Although a similar pattern had been already observed in several nuclear matter studies performed within
G-matrix \cite{Gmatrix,Gmatrix2} and CBF   perturbation theory \cite{FFP}, as well as within the Self Consistent Green's Function approach \cite{SCGF},  the results of these calculations have not been exploited in analyses of the transport coefficients.

In this article, we report the results of a study of the shear viscosity and thermal conductivity coefficients of the fermion hard-sphere 
system, carried out taking into account perturbative contributions up to second order in the CBF effective interaction discussed in I. 
In view of future applications to neutron star matter, the degeneracy of  the momentum eigenstates has been set to $\nu$=2.

In Sec.~\ref{landau} we recollect the expressions of the transport coefficients obtained from the Landau-Boltzmann kinetic equation, while the derivation of the CBF effective interaction for the fermion hard-sphere system with $\nu$=2\textemdash involving additional difficulties with respect to the 
case $\nu$=4\textemdash is described in Sec.~\ref{def:veff}. The calculated effective masses and
in-medium scattering probabilities are discussed in Secs.~\ref{equilibrium} and \ref{sec:scatprob}, respectively, while the shear viscosity and thermal conductivity 
coefficients obtained from our approach are reported in Sec.~\ref{transport}. Finally, In Sec.~\ref{conclusion} we summarize our findings and state the conclusions.  

\section{Transport properties of normal Fermi liquids} 
\label{landau}

In this work, we follow the approach based on Landau's theory of normal Fermi liquids \cite{baym-pethick}, originally 
developed by Abrikosov and Khalatnikov \cite{ak}. Within this scheme, the shear viscosity and thermal conductivity 
coefficients---denoted $\eta$ and $\kappa$, respectively---are determined from the momentum and energy fluxes
obtained from the kinetic equation for the distribution function
\begin{align}
\label{boltzmann}
\frac{ \partial n_{\bf k} }{ \partial t } +  \frac{ \partial n_{\bf k} }{ \partial {\bf r} } \cdot \frac{ \partial e_{\bf k} }{ \partial {\bf k} } 
- \frac{ \partial n_{\bf k} }{ \partial {\bf k } } \cdot \frac{ \partial e_{\bf k} }{ \partial {\bf r}  } = I \left[ n_{\bf k} \right]  \ ,
\end{align}
where $e_{\bf k}$ denotes the energy of a quasiparticle carrying  momentum ${\bf k}$, and $ I \left[ n_{\bf k} \right]$ is the collision integral, 
the definition of which involves the in-medium scattering probability~$W$.

In general, the scattering probability depends on the initial and final momenta of the particles participating in the process. In the low-temperature limit, however, 
the system is strongly degenerate, and only quasiparticles occupying states in the vicinity of the Fermi surface can be involved in interactions.  
As a consequence, the magnitudes of their momenta can be all set equal to the Fermi momentum, and $W$ reduces to a function of the two angular 
variables $\theta$ and $\phi$ only. The former is the angle between the initial momenta,  whereas the latter is the angle between the planes
specified by the initial and final momenta, respectively. 

The above procedure leads to the expressions   \cite{ak}
\begin{align}
\label{eta_AK}
\eta_{AK} = \frac{16}{15}\frac{1}{T^2}\frac{k_F^5}{{m^\star}^4}  \ \frac{1}{ \langle W \rangle (1-\lambda_\eta)} \ ,
\end{align}
and
\begin{align}
\label{cappa_AK}
\kappa_{AK} = \frac{16}{3} \frac{1}{T} \frac{ \pi^2 k_F^3}{{m^\star}^4} \  \frac{1}{\langle W \rangle (3 - \lambda_\kappa)} \ ,
\end{align}
where $T$ is the temperature,  
\begin{align}
\label{lambda:eta}
\lambda_\eta & = \frac{ \langle W [ 1 - 3 \sin^4 (\theta/2) \sin^2 \phi ] \rangle}{\langle W \rangle}  \ , \\
\label{lambda:cappa}
\lambda_\kappa & = \frac{\langle W (1+2 \cos \theta) \rangle}{\langle W \rangle} \ ,
\end{align}
and  the angular average of a generic function $\mathcal{W}(\theta,\phi)$ is defined as
\begin{align}
\label{Wavg}
\langle \mathcal{W} \rangle = \int \frac{d\Omega}{2\pi} \  \frac{ \mathcal{W}(\theta,\phi) }{ \cos (\theta/2)  } \ ,
\end{align}
with $d \Omega = \sin \theta d \theta d \phi$. Note that, as we are considering a system of identical particles, the 
angular integration is normalised to $2 \pi$.

In the above equations, corresponding to the leading terms of low-temperature expansions,  $m^\star$ denotes
the quasiparticle effective mass evaluated at momentum such that $|{\bf k}| = k_F$. Note that the shear viscosity and
thermal conductivity coefficients exhibit different $T$-dependences.

The quasiparticle lifetime $\tau$ can also be written in terms of the angular average  of the scattering probability, Eq. ~\eqref{Wavg},  
according to
\begin{align}
\label{tau_AK}
\tau = \frac{1}{T^2} \ \frac{8\pi^4}{{m^\star}^3}\    \frac{ 1 }{\langle W \rangle} \ .
\end{align}

Corrections to the Abrikosov-Khalatnikov results were derived by Brooker and Sykes in the late 1980 \cite{bs1}. 
Their final results can be cast in the form
\begin{align}
\label{eta_sb}
\eta & = \eta_{AK} \frac{1-\lambda_\eta}{4} \\
\nonumber
& \times \sum_{k=0}^\infty \frac{4k+3}{(k+1)(2k+1)[(k+1)(2k+1)-\lambda_\eta]} \ ,
\end{align}
and 
\begin{align}
\label{cappa_sb}
\kappa & = \kappa_{AK} \frac{3 - \lambda_\kappa}{4} \\ 
\nonumber
& \times \sum_{k=0}^\infty \frac{4k+5}{(k+1)(2k+3)[(k+1)(2k+3)-\lambda_\kappa]} \ .
\end{align}
The effect of the corrections, measured by the ratio between the results of Ref. \cite{bs1} and those of Ref. \cite{ak}, 
while being moderate on viscosity, turns out to be large on thermal conductivity.  One finds $0.750 \leq (\eta/\eta_{AK}) \leq 0.925$, 
 and $0.417 \leq (\kappa/\kappa_{AK}) \leq 0.561$.
 
The above equations show that the input required to obtain $\eta$ and $\kappa$ includes the effective masses,  the calculation of which has been discussed in I, and the in-medium scattering probability, which 
 can be  obtained in Born approximation using the  CBF effective interaction. 

\section{CBF effective interaction}
\label{def:veff}

The fermion hard-sphere system is a collection of spin-one-half particles, the dynamics of which 
are described by the Hamiltonian
\begin{align}
\label{def:H}
H = \sum_{i} \frac{{\bf k}_i^2}{2m} + \sum_{j>i} v(r_{ij})  \ , 
\end{align}  
where ${\bf k}_i$ is the momentum of the $i$-th particle, $r_{ij} = |{\bf r}_i - {\bf r}_j|$ denotes the distance between particles $i$ and $j$ and
\begin{equation}
\label{potential}
v(r) = \left\{
\begin{array}{cc}
\infty \ \ ,  & \ \ r < a \\
0    \ \ \  ,  &  \ \ r > a 
\end{array}
\right. \ .
\end{equation}
By setting the degeneracy of the momentum eigenstates to $\nu$=2 or 4 
this system can be regarded as a model of neutron matter or isospin-symmetric nuclear matter, respectively \cite{FW}. 
It should be kept in mind,  however, that this analogy is limited to densities below the solidification point.
 The analysis of Ref.~\cite{solidification} indicates that for $\nu$=2 solidification occurs at a density $\rho_s$ such that
${\rho}_s a^3$=0.23.  Moreover, at subnuclear densities nuclear matter is known to undergo transitions to superfluid and/or
superconducting phases, which are not allowed by the purely repulsive interaction of Eq.~~\eqref{potential}.  
 Theoretical studies suggest that neutron matter becomes superfluid at density $\lsim$0.08 fm$^{-3}$ \cite{superfluid}. 

The results reported in the following sections have been obtained setting $m$=1 fm$^{-1}$ and $a$=1 fm.

\subsection{Definition of the CBF effective interaction}

The CBF effective interaction, whose detailed derivation can be found in I, is defined as \cite{shannon,BV}
 \begin{equation}
\label{veff:final2}
v_{\rm eff}(r) = \frac{1}{m} \left[ {\boldsymbol \nabla f(r)} \right]^2 \ , 
\end{equation}
where the function $f(r)$, describes the correlation structure induced by the potential of Eq.~~\eqref{potential}.  
The shape of $f(r)$ is obtained from functional minimization of the expectation value of the  Hamiltonian ~\eqref{def:H}
in the {\em correlated} ground state, defined as 
\begin{equation}
\label{CBF:GS}
|  \Psi_0 \rangle     =    \frac{ F |  \Phi_0 \rangle } {  \sqrt{ \langle  \Phi_0 | F^{\dagger} F | \Phi_0 \rangle } } \ .
\end{equation}
In the above equation, 
\begin{equation}
F = \prod_{j>i} f(r_{ij}) \ , 
\end{equation}
while the state  
$ | \Phi_0 \rangle$,  describing the system in the absence of interactions,
is represented by a Slater determinant of single particle wave functions,  consisting of a plane wave and Pauli spinors
associated with  the discrete degrees of freedom. For any fixed density $\rho$, all momentum eigenstates corresponding to $|{\bf k}| < k_F$, 
with $k_F = (6 \pi^2 \rho/\nu)^{1/3}$, are occupied with unit probability.

Within the  two-body cluster approximation (see, e.g., Ref.~\cite{jwc}), this procedure yields an Euler\textendash Lagrange equation, to be solved with the boundary conditions dictated by the properties of the hard-core potential,  as well as by the requirement that correlation effects vanish for large separation distances
\begin{align}
\label{eq:boundaryfa}
f(r \le a) = 0  \ \ \ , \ \ \ f( r \geq d ) = 1  \ .
\end{align}
The additional constraint
\begin{align}
\label{eq:boundaryfprime}
f^{\prime} (d)  = 0 \ ,
\end{align}
that can be fulfilled introducing a Lagrange multiplier $\lambda$,
enforces continuity of the logarithmic derivative of the two-particle wave function at $r = d$.

The explicit form of the Euler\textendash Lagrange equation  is
\begin{equation}
\label{ELeq}
g^{\prime \prime}(r) - g(r) \left[ \frac{\Phi^{\prime \prime}(r)}{\Phi(r)}  + m \lambda \right] = 0 \ , 
\end{equation}
where
\begin{align}
\label{eq:grel}
g(r) = f(r) \Phi(r) \ ,
\end{align}
with
\begin{align}
\label{eq:phiel}
\Phi(r)   \equiv r \sqrt{1 - \frac{1}{\nu} \ell^2 ( k_F r) } \ . 
\end{align}
In the above equation, $\ell(x) = 3(\sin x - x \cos x)/x^3$ is the Slater function describing the effects of statistical correlations.

For any given values of density, $\rho$,  and correlation range, $d$,
Eq.~\eqref{ELeq} can be solved numerically to obtain the  correlation function $f(r)$,  with the lagrange multiplier $\lambda $ determined by 
the requirement that Eq.~\eqref{eq:boundaryfprime} be satisfied. The correlation range $d $ is treated as a free parameter, 
to be adjusted in such a way as to have
\begin{align}
\label{newdef}
E_0   = \frac{3}{5} \frac{k_F^2}{2m} + \frac{\rho}{2} \int_a^d d^3 r \ v_{\rm eff}(r) \left[ 1 - \frac{1}{\nu} \ell^2(k_F r) \right] \ , 
\end{align}
where $E_0 = \langle H \rangle / N$ is the ground-state expectation value  of the Hamiltonian computed using some advanced and accurate many-body technique, 
such as the variational Fermi Hyper-Netted Chain (FHNC) approach employed in I for the hard-sphere system of degeneracy $\nu$=4. 
Note that the quantity appearing on the right hand side of Eq.~\eqref{newdef} is the expectation value of the effective Hamiltonian
\begin{align}
H_{\rm eff} = \sum_{i} \frac{{\bf k}_i^2}{2m} + \sum_{j>i} v_{\rm eff}(r_{ij})  \ , 
\end{align}
in the Fermi gas ground state.

From Eqs.~~\eqref{ELeq}-~\eqref{eq:phiel} 
it clearly appears that $f(r)$  depends explicitly on the degeneracy of the system, $\nu$,  through the  coefficient of the squared Slater function. The 
lower the degeneracy of the system, the larger the effect of these correlations, the range
of which monotonically increases as the density decreases. As a consequence, in the low-density region the 
determination of $f(r)$ from the numerical solution of Eq. ~\eqref{ELeq} with $\nu$=2 turns out to be hindered by the presence of long-range  statistical correlations, whose effect is significantly larger than in the case $\nu$=4.

Owing to the above difficulty, at $c=k_F a <  0.5$, the  value of $E_0$ obtained within the FHNC approach does not develop a clear minimum as
a function of the correlation range, which in this case plays the role of variational parameter. Therefore, it does not provide an upper bound to the ground-state energy of the system. 
To overcome this problem, and obtain the accurate estimate of $E_0$
needed to determine the CBF effective interaction at all densities, we have employed the ground-state expectation values of
the Hamiltonian of Eq.~~\eqref{def:H} obtained from both Variational Monte Carlo  (VMC) and Diffusion Monte Carlo (DMC) calculations. 

\subsection{Monte Carlo calculation of $\langle H \rangle/N$}

Within VMC, the multidimensional integrations involved in the calculation of the expectation value of the Hamiltonian in the correlated ground state  are performed using Metropolis Monte Carlo quadrature \cite{metropolis:1953}. The trial wave function, chosen to be the same  as in the FHNC calculation, is defined as
\begin{equation}
\label{trialwf}
\Psi_T(\mathbf{R})=  \langle \mathbf{R} | \Psi_0  \rangle \ , 
\end{equation}
where $ | \Psi_0 \rangle $ is given by Eq. ~\eqref{CBF:GS}, $\mathbf{R} \equiv \{{\bf r}_1, \ldots , {\bf r}_N \}$ denotes the set of coordinates specifying the system in configuration space,
and $| \mathbf{R} \rangle$ is the corresponding eigenstate.

The infinite system is modeled by considering a finite number of particles in a box, and imposing periodic boundary conditions. As a consequence,
the spectrum of eigenvalues of  the momentum  $\mathbf{k}$ is discretized. For a cubic box of side $L$, one finds the familiar result
\begin{equation}
k_i=\frac{2\pi}{L} n_i \ \ , \ \  i=x,y,z  \ \ , \ \  n_i=0,\pm 1, \pm 2, \dots \ .
\end{equation}
In order for the  wave function to describe a system with vanishing total momentum and angular momentum, all shells corresponding to momenta
such that $|{\bf k}| < k_F$ must be filled. This requirement determines a set of ``magic numbers'', which are commonly employed in simulations
of periodic systems. For example, the VMC\textemdash as well as DMC\textemdash calculations whose results  are reported in this article  have been performed with 132 particles, corresponding to 66 and 33 momentum eigenstates for degeneracy $\nu$=2 and 4, respectively.

The expectation value of the Hamiltonian in the state described by the trial wave function of Eq. ~\eqref{trialwf} can be cast in the form
\begin{equation}
\langle \Psi_T | H | \Psi_T\rangle = \int d\mathbf{R} \  E_L(\mathbf{R}) P(\mathbf{R}) \ ,
\end{equation}
where the local energy $E_L(\mathbf{R}) $ is defined as
\begin{equation}
E_L(\mathbf{R}) = \frac{H\Psi_T(\mathbf{R})}{\Psi_T(\mathbf{R})} \ ,
\end{equation}
and we have introduced the probability density $P(\mathbf{R})~\equiv~\Psi_{T}^*(\mathbf{R}) \Psi_T(\mathbf{R})$. Within VMC, the above integral
is estimated by a sum over the set $ \{{\bf R} \} $, consisting of $N_c$ configurations  sampled from the distribution $P(\mathbf{R})$ using the Metropolis algorithm
\begin{equation}
\langle \Psi_T |  H | \Psi_T \rangle \approx \frac{1}{N_c} \  \sum_{{\bf R}_i \in \{\mathbf{R}\}} E_L(\mathbf{R}_i)   \ .
\end{equation}

The VMC approach can be seen as an alternative to the cluster expansion technique underlying the FHNC approach, allowing for a stringent test of the approximation implied by the  neglect of  cluster contributions associated with the so-called elementary diagrams \cite{jwc}.

The main drawback of VMC, obviously shared by FHNC,
is that the accuracy of the result entirely depends on the quality of the trial wave function. The DMC method \cite{kalos:2012,grimm:1971} overcomes the limitations of the variational approach by using a projection technique to enhance the true ground-state component of the trial wave function. This
result is achieved expanding $|\Psi_T\rangle$  in eigenstates of the Hamiltonian according to
\begin{equation}
|\Psi_T\rangle=\sum_n c_n |n\rangle \ \ , \ \  {H}|n\rangle = E_n |n\rangle\ ,
\end{equation}
which implies
\begin{equation}
\lim_{\tau\to\infty}e^{-({H}-E_0)\tau} |\Psi_T\rangle=c_0 |0\rangle\, ,
\end{equation}
with $\tau$ being the imaginary time. Provided  $|\Psi_T \rangle$ it is not
orthogonal to the true ground state, i.e. for $c_0\neq 0$, in the limit of large $\tau$ the above procedure projects out the exact lowest-energy state.

Because the direct calculation of $\exp[-({H} -E_0)\tau]$ involves prohibitive difficulties, the imaginary-time evolution is decomposed into $N$ small imaginary-time steps,  and complete sets of position eigenstates are inserted, in such a way that only the calculation of the short-time propagator is required. This procedure yields the expression
\begin{align}
\label{eq:propagation}
& \langle \mathbf{R}_{N+1}   | e^{-({H}-E_0)\tau}  | \mathbf{R}_1 \rangle \\
\nonumber
& = \int \  d{\bf R}_2 \ldots d{\bf R}_N \  \langle \mathbf{R}_{N+1} | e^{-({H} -E_0)\Delta\tau} | \mathbf{R}_{N}\rangle \\
 & \times \langle \mathbf{R}_{N} | e^{-({H} -E_0)\Delta\tau} | \mathbf{R}_{N-1} \rangle \dots \nonumber 
  \langle \mathbf{R}_2| e^{-({H} -E_0) \Delta \tau} | \mathbf{R}_1 \rangle  \ , 
\end{align}
where, for the sake of simplicity, the dependence on the discrete  degrees of freedom has been omitted. Monte Carlo techniques are used to sample the
paths $\mathbf{R}_i$ in the propagation. Note that, although Eq.~~\eqref{eq:propagation} is only exact in the  $\Delta\tau\to 0$ limit, its accuracy 
can be tested performing several simulations with smaller and smaller time step and extrapolating to zero.

\begin{figure}[ht]
\begin{center}
\includegraphics[width= 0.425\textwidth]{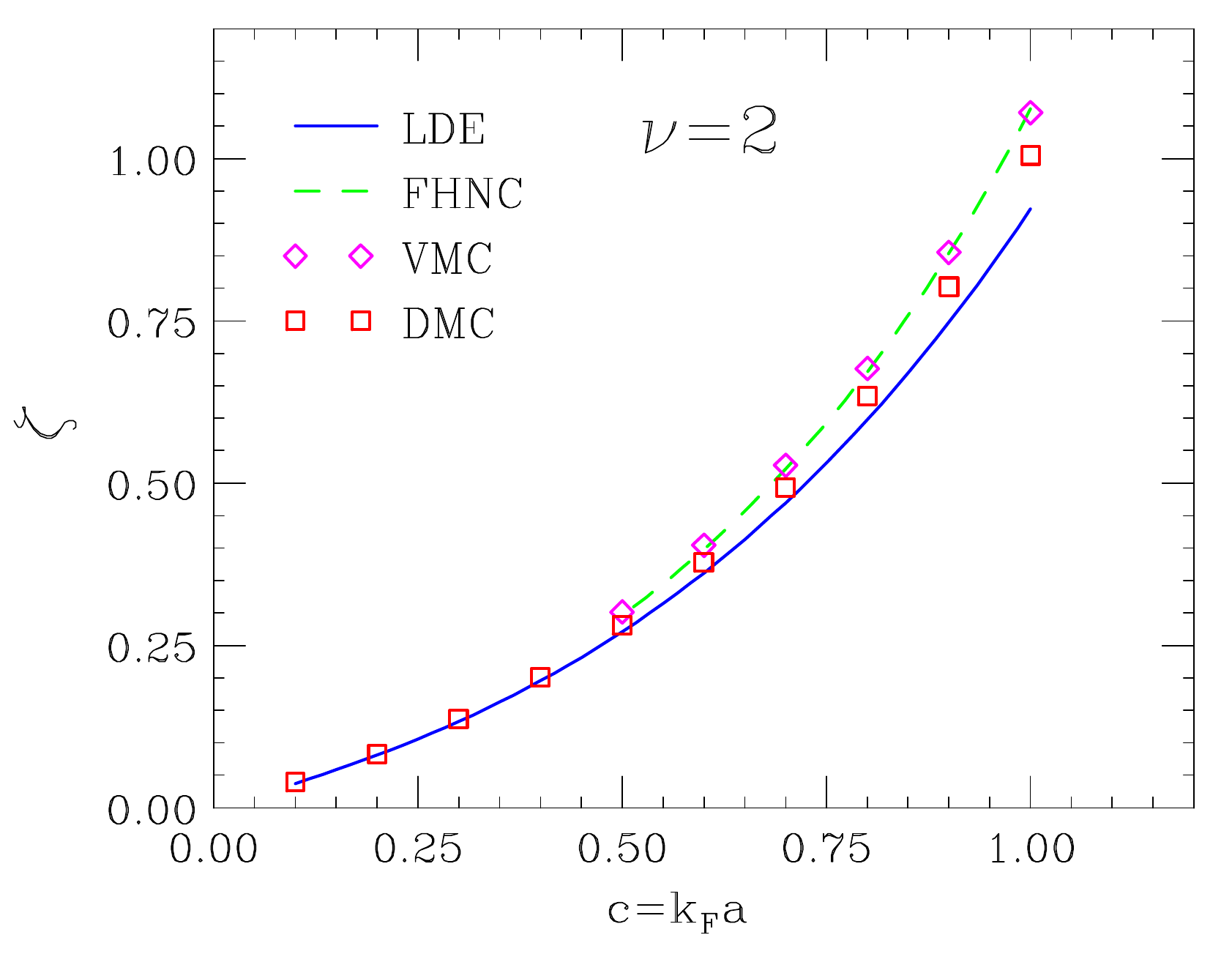}
\caption{(color online) c-dependence of the quantity $\zeta$, defined by Eq. ~\eqref{def:z}, for degeneracy $\nu$=2. The solid and dashed lines show 
the results obtained from the low-density expansion and the variational 
FHNC approach, respectively. The VMC and DMC results are represented by diamonds and squares. Monte Carlo error 
bars are not visible on the scale of the figure.\label{fig:MC2}}
\end{center}
\end{figure}

\begin{figure}[ht]
\begin{center}
\includegraphics[width= 0.425\textwidth]{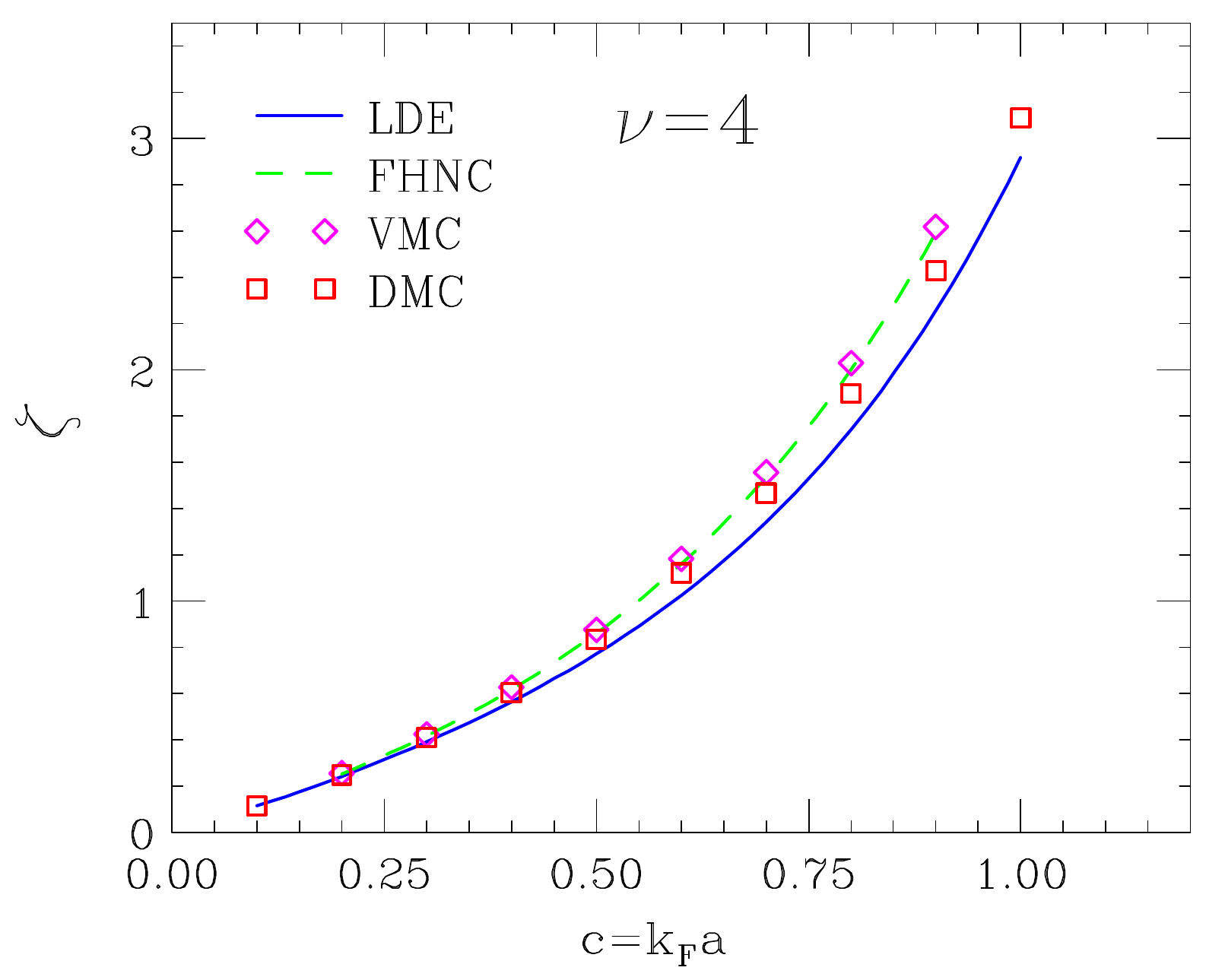}
\caption{(color online) Same as in Fig. \ref{fig:MC2}, but for degeneracy $\nu$=4.\label{fig:MC4}}
\end{center}
\end{figure}

The ground-state  energy  can be  conveniently written in terms of the dimensionless quantity $\zeta$, parametrizing the deviation from 
the energy of the non interacting system, defined through the equation
\begin{equation}
\label{def:z}
E_0 = \frac{3}{5} \frac{k_F^2}{2m} \left( 1 + \zeta \right) \ .
\end{equation}
In Figs. \ref{fig:MC2} and \ref{fig:MC4} the results of DMC calculations of  $\zeta$ are compared to the values obtained from the VMC and FHNC approaches\textemdash the latter being only available at $c~\geq~0.5$\textemdash as well as to the predictions of the low-energy expansions, obtained from  \cite{bishop}
\begin{align}
\label{eq:E0nu2}
E_0  =  \frac{k_F^2}{2 m}   \bigg[   \ \frac{3}{5}    & +   \frac{2}{3 \pi } c +   \frac{4}{35 \pi ^ 2 }      \left (11 - 2 \log 2  \right)c^2  \bigg. \\
\nonumber
& \bigg. + 0.230 c^3   \ \bigg]  \  , 
\end{align}
 for $\nu$=2 and
 \begin{align}
 \label{eq:E0nu4}
E_0 =  \frac{k_F^2}{2 m} & \left[  \  \frac{3}{5}   +     \frac{2}{ \pi} c   +    \frac{12}{35 \pi ^ 2 }    \left( 11 - 2 \log 2  \right) c ^ 2   \right. \\
\nonumber
 &   \left.  \  +  \  0.780 c^3  +  \frac{32}{9 \pi^3} \left( 4     \pi - 3 \sqrt{3}  \right)  c^4 \log  c   \  \right]  \ , 
 \end{align}
 for $\nu$=4.  
 
It clearly appears that the VMC and FHNC
results are very close to one another,  thus showing that at $c\geq 0.5$ the FHNC approximation does provide an upper bound to the ground-state energy.
The accuracy of the variational result is measured by the difference between the VMC\textemdash or, equivalently, FHNC\textemdash values of $\zeta$ and those obtained from DMC. In the case of degeneracy $\nu$=2, illustrated in Fig. \ref{fig:MC2}, this difference ranges between  
$\sim$2\% and $\sim$9\% at  $0.2 \leq c \leq 1$. Note that a 9\% difference in $\zeta$ translates in a difference of less that 3\% in the 
ground-state energy $E_0$. The low-density expansion turns out to be also quite accurate, its predictions being within
5\% of the DMC results at $c < 0.5$. Figure  \ref{fig:MC4} shows the results corresponding to $\nu=4$, which exhibit the same
pattern.


To summarize, the CBF effective interaction of the hard-sphere system with $\nu$=2 has been computed from Eq. ~\eqref{veff:final2}, choosing the
correlation range $d$ in such a way as to reproduce the ground-state expectation value of the Hamiltonian, $E_0$, obtained using the
DMC technique.

\section{Quasiparticle spectrum and effective mass}
\label{equilibrium}

Once the effective interaction has been determined, the calculation of the quasiparticle energy spectrum and effective mass can be performed 
following the procedure described in I for the case of degeneracy $\nu$=4.

The resulting spectrum
\begin{align}
\label{def:spectrum}
e(k) = e_0(k) + {\rm Re} \  \Sigma[k,e(k)] \ ,
\end{align}
has been computed including the first and second order contributions to the proper self-energy $\Sigma$. The former provides the 
energy-independent correction corresponding to the Hatree-Fock approximation, while the latter brings about an explicit energy dependence. 

\begin{figure}[htbp]
\begin{center}
\includegraphics[width= 0.45\textwidth]{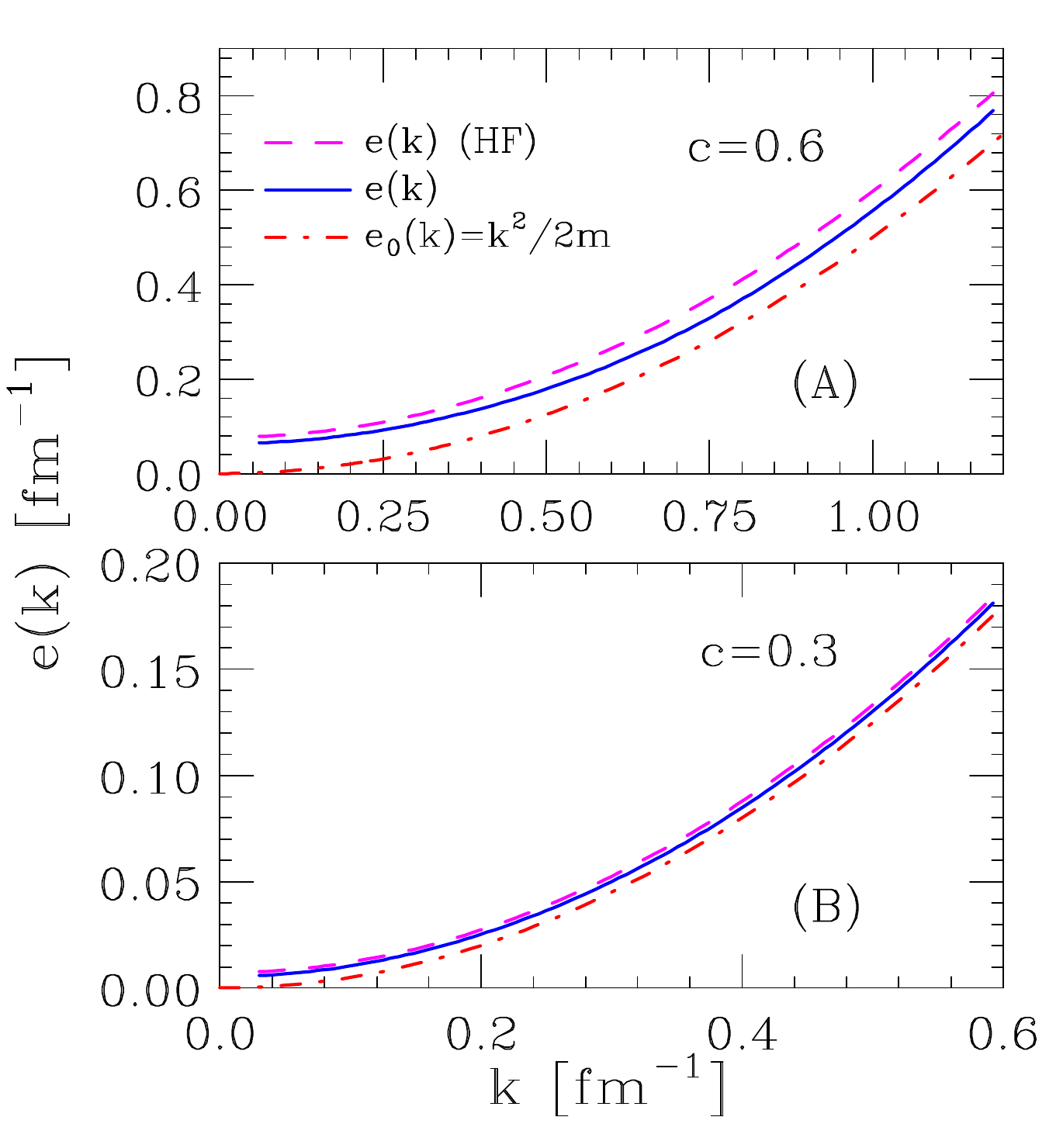}
\caption{(color online) Quasiparticle energy of the fermion hard-sphere system of degeneracy $\nu$=2 at $c=0.3$ [panel (B)]
and 0.6 [panel(A)].  The dashed and solid lines correspond to the first order (\emph{i.e.} Hartree-Fock) and second order
approximations to the proper self-energy $\Sigma$, respectively. For comparison, the dot-dash lines show the kinetic energy spectrum. }
\label{fig:spe_nu2}
\end{center}
\end{figure}

\begin{figure}[htbp]
\begin{center}
\includegraphics[width= 0.425\textwidth]{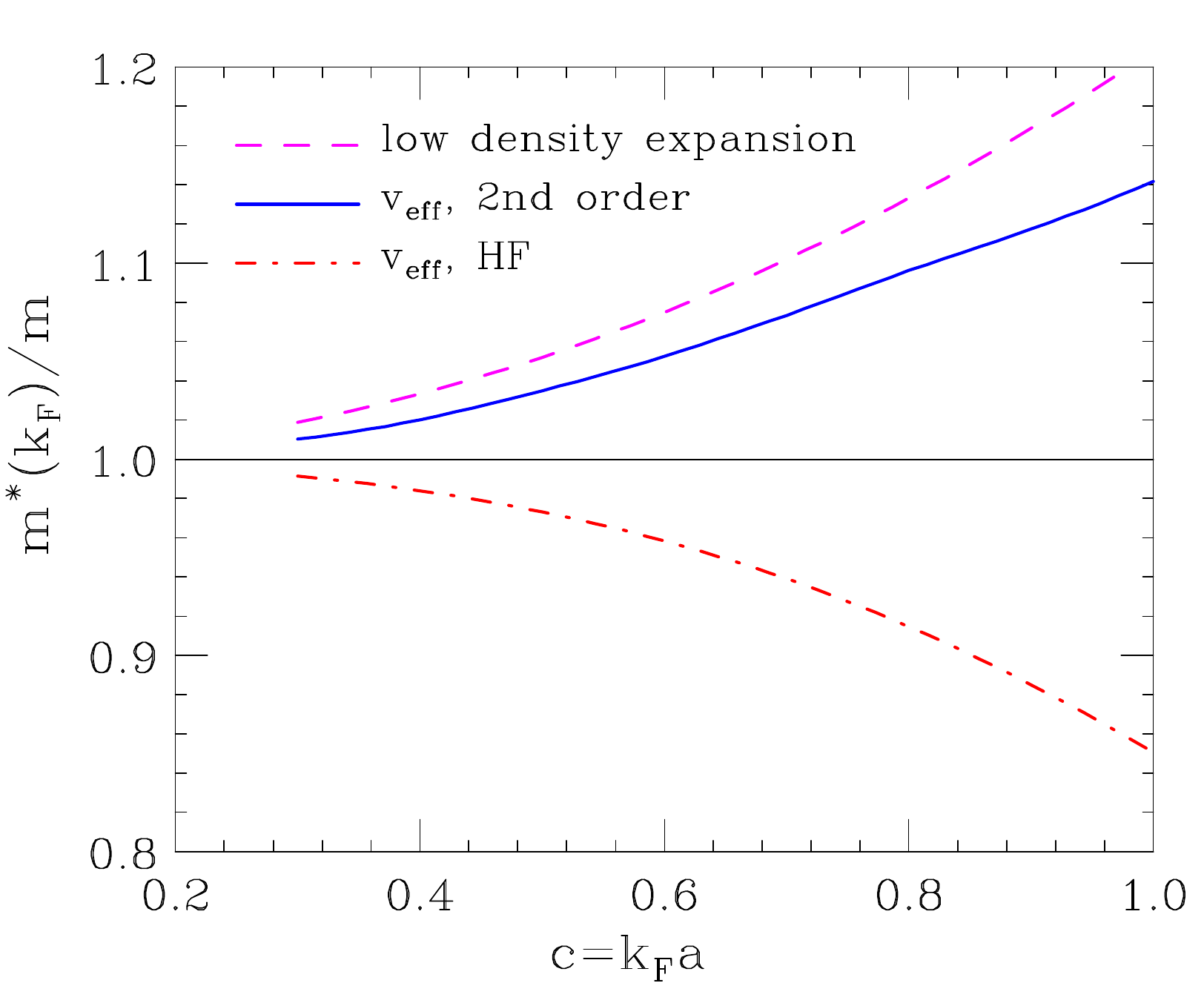}
\caption{(color online) $c$-dependence of the ratio $m^\star(k_F)/m$ for the hard-sphere system of
degeneracy  $\nu$=2. The dot-dash and solid lines represent the results of calculations carried out using the first and second order approximations to the self-energy.  For comparison, the dashed line shows the results computed using the low-density expansion of Eq.~~\eqref{pert_mstar}.
}
\label{fig:mstar_nu2}
\end{center}
\end{figure}

The momentum dependence of the quasiparticle energies at $c=$0.3 and 0.6 is displayed in Fig. \ref{fig:spe_nu2}, while 
 Fig.~\ref{fig:mstar_nu2} shows the corresponding effective mass, obtained  from 
\begin{align}
\label{def:mstar}
m^\star(k) =  \left( \frac{1}{k} \frac{ d e}{d k} \right)^{-1} \ .
\end{align}

Second order corrections to the self-energy have the same effects  observed in I for the case $\nu=4$.
The appearance of the energy dependence results in small modifications of the quasiparticle spectrum, but  dramatically  affects
bot the magnitude and the density dependence of the effective mass at $|{\bf k}| = k_F$.

For reference, Fig.~\ref{fig:mstar_nu2} also shows the effective mass computed using the low-density expansions \cite{galitskii}
\begin{align}
\label{pert_mstar}
\frac{m^\star(k_F)}{m} = 1 + \frac{ 8 }{ 15 \pi^2 }( 7 \ln 2 - 1 ) c^2 \ . 
\end{align}
The difference between the value of  $m^{\star}(k_F)$ evaluated using the  CBF effective interaction and the one obtained  
from Eq. ~\eqref{pert_mstar} turns out to be $\leq 2\%$ for $c\leq 0.6$, and grows up to 5.5\% as the value of $c$ increases up to $c=1.0$.

\section{Scattering probability}
\label{sec:scatprob}

The scattering probability $W(\theta, \phi)$ appearing in the collision integral of Eq.~~\eqref{boltzmann} is trivially related to  the scattering
amplitude $f(\theta, \phi)$ through the relation
\begin{align}
\label{eq:W}
W(\theta, \phi) = 
\pi \left |  \frac{4 \pi}{ m }  f(\theta, \phi) \right |^2   \ .
\end{align}
The scattering amplitude is in turn related to  the differential cross section according to
\begin{align}
\label{eq:scat_ampl}
\frac{d \sigma}{ d \Omega } = |f(\theta,\phi) |^2 \ . 
\end{align}
Combining the above equations one finds\footnote{Note that the relation between scattering probability and cross section reported in Ref. \cite{BV}, in which the
factor $16 \pi^3/m^2$  is replaced by $16 \pi^2/m^2$,  is incorrect.}
\begin{align}
\label{dsigma:W}
W(\theta, \phi) = \frac{16      \pi^3}{m^2}  \left( \frac{d\sigma}{d\Omega} \right) \ .
\end{align}

The scattering cross section is usually expressed in either the laboratory (L) or the center-of-mass (CM) frame. However, the
Abrikosov\textendash Khalatnikov formalism is derived in a different  frame, referred to as AK frame, in which the
Fermi sphere is at rest.

To clarify the connection between AK and CM frames, let us consider the process in which two particles carrying momenta ${\bf k}_1$ and
${\bf k}_2$ scatter to final states of momenta ${\bf k}_1^\prime$ and ${\bf k}_2^\prime$.
The total energy of the initial state
\begin{align}
E = \frac{{\bf k}_1^2}{2m} + \frac{{\bf k}_2^2}{2m} \ ,
\end{align}
can be conveniently rewritten in terms of the center of mass and relative momenta,
${\bf K}~= {\bf k}_1~+~{\bf k}_2$ and ${\bf k}~=~({\bf k}_1~-~{\bf k}_2)/2$, as
\begin{align}
E = \frac{{\bf K}^2}{2M} + \frac{{\bf k}^2}{2\mu} = {\cal E}+ {\cal E}_{{\rm rel}} \ ,
\end{align}
with $M = 2m$ and $\mu = m/2$. In the CM reference frame, in which the center of mass of the system is at rest,
$E~=~E_{{\rm cm}}~=~{\cal E}_{{\rm rel}}$, while in the L frame, in which ${\bf k}_2=0$,
$E=E_{{\rm L}}= 2{\cal E}_{{\rm rel}}$.

In strongly degenerate systems, the magnitude of all momenta playing a role in the determination of the transport coefficients is equal to the Fermi momentum, and
conservation of energy requires that the angle between the
momenta of the particles participating in the scattering process be the same before
and after the collision. In general, however, the angle $\phi$ between the initial and
final relative momenta, ${\bf k}$ and ${\bf k}^\prime = ({\bf k}_1^\prime - {\bf k}_2^\prime)/2$, defined
through
\begin{align}
\cos \phi = \frac{ ({\bf k} \cdot {\bf k}^\prime) }{ |{\bf k}||{\bf k}^\prime| } \ ,
\label{def:phi}
\end{align}
does not vanish. Hence, for any given Fermi momentum, i.e. for any given matter density, the
scattering process in the AK frame is specified by the center of mass energy
\begin{align}
{\cal E}^{{\rm AK}} = \frac{k_F^2}{2m} (1 + \cos \theta ) \ ,
\end{align}
and the two angles $\theta$ and $\phi$.

The AK-frame variables can be easily connected to those of the  CM reference frame. Since the relative kinetic energy, \emph{i.e.} the energy in the CM reference frame, $E_{\rm cm}$,  is the same in any frame, we have
\begin{align}
\label{ecm}
E_{\rm  cm} = {\cal E}^{ AK}_{\rm  rel}  = \frac{k_F^2}{2m} (1-\cos \theta)  \ , 
\end{align}
where we have used again the condition that  scattering processes involve particles in momentum states close to the Fermi surface.
Moreover, the  angle between the planes containing ingoing and outgoing momenta, $\phi$ , is nothing but the angle  between the initial and final relative  momenta, and can therefore be identified with the scattering angle in the CM frame, setting
\begin{align}
\label{thetacm}
\Theta_{\rm cm} = \phi \ .
\end{align}
Through the above relations, the differential cross section in the CM frame, written as a function of the two variables $E_{\rm  cm}$ and
$\Theta_{\rm cm}$, can be transformed into the corresponding quantity in the AK frame, depending on the two angular variables
$\theta$ and $\phi$, needed for the calculation of the transport coefficients. We can write
\begin{equation}
\frac{d \sigma}{ d \Omega} [ E_{\rm cm}(\theta),  \Theta_{\rm cm}(\phi)] =  \frac{d \sigma}{ d \Omega} ( \theta,  \phi ) \ , 
\end{equation}
with $E_{\rm cm}(\theta)$ and $\Theta_{\rm cm}(\phi)$ given by Eqs. ~\eqref{ecm} and ~\eqref{thetacm}.

In the pioneering work of Ref. \cite{transpII}, the scattering probability in neutron star matter was computed from Eq.~\eqref{dsigma:W} replacing the
bare nucleon mass with an effective mass and using the nucleon-nucleon scattering cross section in free space, obtained from the measured phase shifts.
This procedure accounts for the fact that both the incoming flux and the phase space available to the final state particles are affected by the presence of the medium. However, it neglects possible medium modifications of the scattering probability.

The authors of Ref. \cite{BV} improved upon the approximation of Ref. \cite{transpII}, using the CBF effective interaction to obtain both the effective mass and the in-medium scattering cross section of pure neutron matter within a consistent framework.

In this work, we apply the approach of Ref. \cite{BV} to the fermion  hard-sphere system. The in-medium scattering probability
has been computed in Born approximation using  the CBF effective interaction and the definition
\begin{align}
W(\theta, \phi ) = \pi  \left| \langle  {\bf k}_1^{\prime},{\bf k}_2^{\prime} |  v_{\rm eff}  |   {\bf k}_1 ,  {\bf k}_2  \rangle \right|^2  \ ,
\end{align}
where ${\bf k}_i$ and ${\bf k}_i^{\prime}$ are the  initial and final momenta, respectively.
The calculation of the matrix element is essentially the same as that performed in I to obtain the second order contributions to the proper self-energy.
The only difference stems from the fact that, because we are considering a scattering process,
here we need to average over the spins of the initial state particles. The result can be written in the form
\begin{align}
 \frac{1}{\nu^2}\sum_{\sigma, \sigma^{\prime}  } & 
\left | \langle {\bf k}_1^{\prime},{\bf k}_2^{\prime} |  v_{\rm eff}  |   {\bf k}_1 ,  {\bf k}_2  \rangle \right|^2  \\
\nonumber
& = M^2 (u)     +  M^2(v)    -  \frac{2}{\nu}  M(u) M(v)  \  ,
\end{align}
where $ M(u)$ and $M(v)$ denote the Fourier transforms of the effective potential evaluated at 
\begin{align}
 u &= | {\bf k} - {\bf k}^{\prime} |  =    k_F\sqrt{ (1 - \cos \theta) (1 - \cos \phi )  }  \ , \nonumber \\
 v &= | {\bf k} + {\bf k}^{\prime} | =    k_F\sqrt{ (1 - \cos \theta) (1 + \cos \phi )  }   \ . 
 \end{align}

\begin{figure}[htbp]
\begin{center}
\includegraphics[width= 0.425\textwidth]{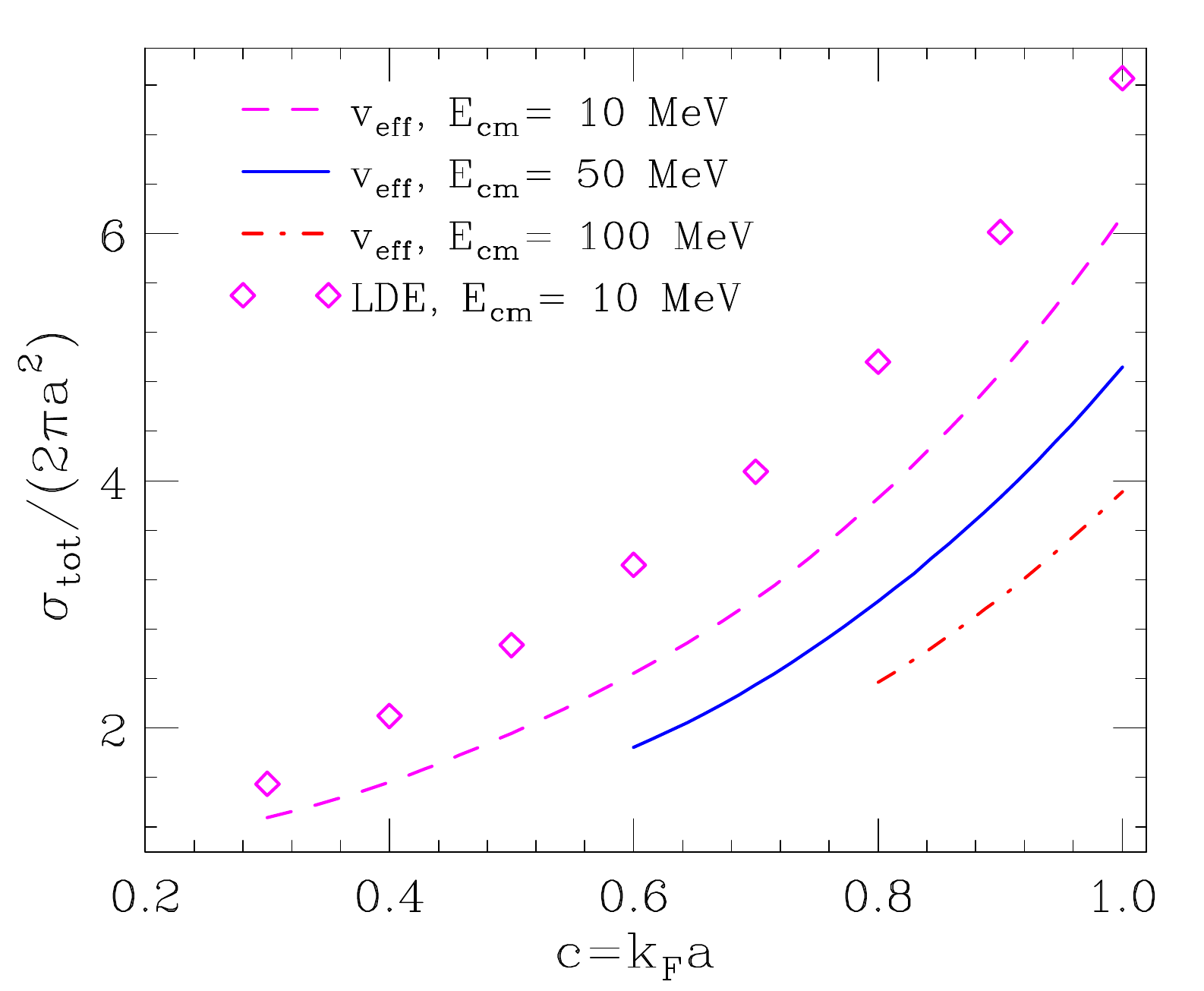}
\caption{(color online) $c$-dependence of the in-medium total cross section of the
fermion hard-sphere system with $\nu$=2\textemdash normalized to the low-energy limit in vacuum\textemdash computed using the CBF effective interaction for different values of the CM energy $E_{\rm cm}$. For comparison , the diamonds show the results of the low-density expansion 
derived in Ref.~\cite{JLTP1}.}
\label{fig:sigma_E}
\end{center}
\end{figure}

\begin{figure}[htbp]
\begin{center}
\includegraphics[width= 0.425\textwidth]{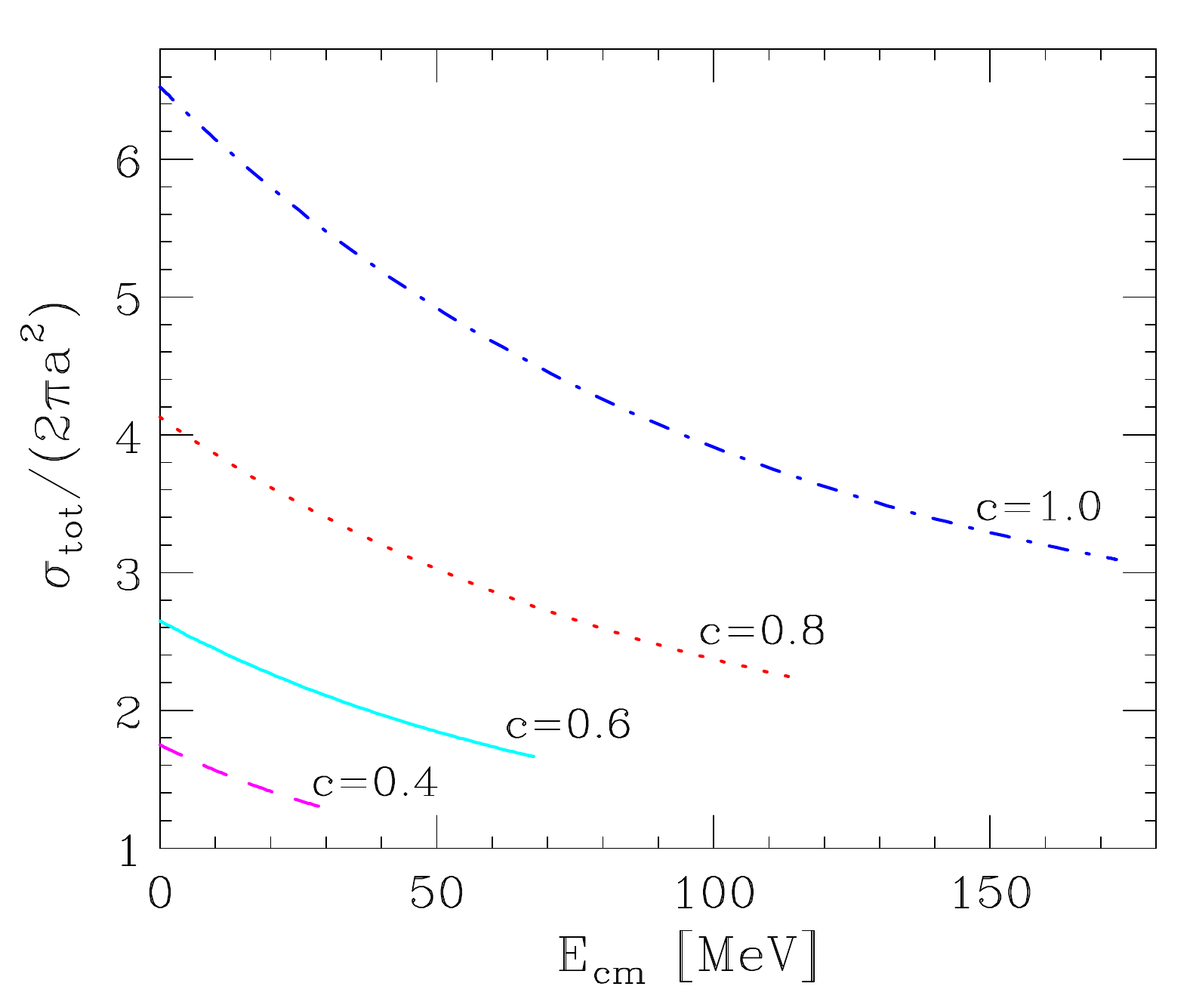}
\caption{(color online) $E_{\rm cm}$-dependence of the in-medium total cross sections  of the
fermion hard-sphere system with $\nu$=2\textemdash normalized to the low-energy limit in vacuum\textemdash computed using the CBF effective interaction for different values of $c = k_F a$.}
\label{fig:sigmatot}
\end{center}
\end{figure}

 The density dependence of the total cross section
 \begin{align}
 \label{sigmatot}
 \sigma_{\rm tot} = \int d \Omega \ \left(  \frac{d\sigma}{d\Omega} \right) , 
 \end{align}
 resulting from our calculations is shown in Fig.~\ref{fig:sigma_E} for center of mass energies $0 \leq E_{\rm cm} \leq 140$ MeV. For any given value 
 of $E_{\rm cm}$, Eq. ~~\eqref{ecm} implies that the Fermi momentum must satisfy  the constraint $k_F > \sqrt{ m E_{\rm cm} }$. Note that 
 $\sigma_{\rm tot}$ is normalized to the low-energy limit obtained from the partial wave expansion of the cross section in vacuum, $\sigma_{\rm tot} = 2 \pi a^2$.
 In Fig.~\ref{fig:sigmatot}, the same quantity is  shown as a function of CM energy, with $E_{\rm cm} < k_F^2/m$,  for different  densities 
 in the range  $0.4 \leq c \leq 1$.


The in-medium scattering probability, defined as in  Eq.~~\eqref{eq:W},
has been also studied within the framework of
standard perturbation theory  \cite{JLTP1,JLTP2}. The authors of Ref. \cite{JLTP1} were able to obtain  the expression of $W(\theta, \phi)$  by solving the  generalised  Bethe\textendash Salpeter  equation for the scattering amplitude of a dilute gas of Fermi hard spheres including terms up to 
order $c$, corresponding to order $c^2$ for the scattering probability. For comparison, in Fig.~\ref{fig:sigma_E} their results at $E_{\rm cm}=10$ MeV are shown by the diamonds. 

The results displayed in Figs.~\ref{fig:sigma_E} and \ref{fig:sigmatot}, showing that $\sigma_{\rm tot}$ increases with density,  can be 
explained considering that the range of  the effective interaction\textemdash that takes into account screening arising form dynamical correlations\textemdash  is larger than the hard-sphere radius $a$, and grows with $c$ (see Fig. 4 of I). 

\section{Transport coefficients}
\label{transport}

\begin{figure}[htbp]
\begin{center}
\includegraphics[width= 0.425\textwidth]{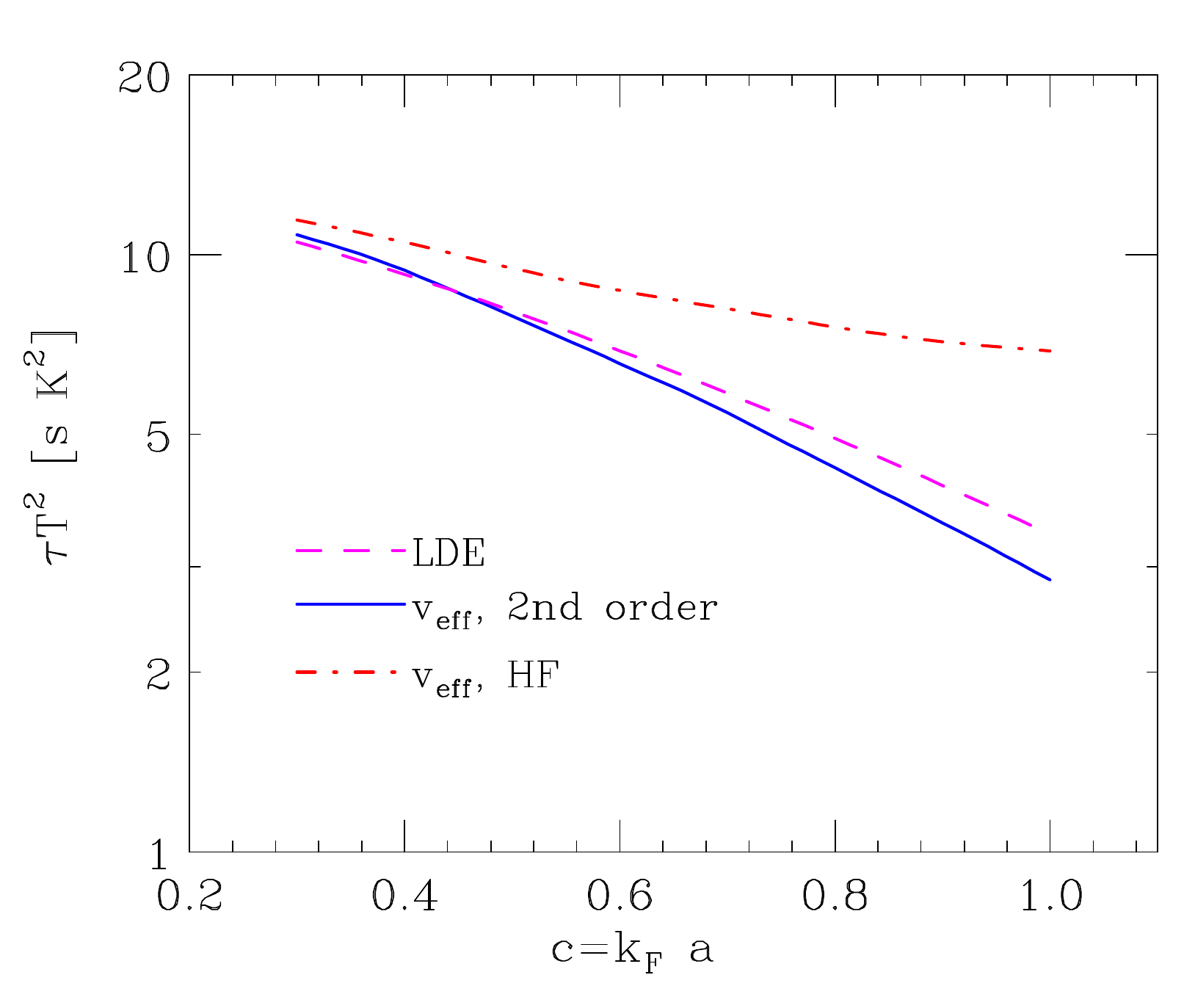}
\caption{(color online) $c$-dependence of the temperature-independent quantity $\tau T^2$, where
$\tau$ is the quasiparticle lifetime of the fermion hard-sphere system of degeneracy  $\nu$=2. The dot-dash and solid lines represent the results of calculations carried out using the first and second order approximations for the effective mass, respectively, while the dashed line shows the results of the low-density expansion of Ref. \cite{JLTP2}, including terms of order up to $c$.}
\label{lifetime}
\end{center}
\end{figure}

The effective mass and scattering probability discussed in the previous sections have been employed to obtain the transport coefficients of the fermion hard-sphere system
with $\nu$=2. Before analyzing  the shear viscosity and thermal conductivity, in Fig. \ref{lifetime} we illustrate the $c$-dependence of the 
temperature-independent quantity
$\tau T^2$, where $\tau$  is the  quasiparticle lifetime of Eq. ~\eqref{tau_AK} computed using the CBF effective interaction.
For comparison the prediction of the low-density expansion of Ref. \cite{JLTP2} is also shown. Overall, the emerging pattern reflects the one observed
in Fig.~\ref{fig:mstar_nu2}. As expected, the large corrections to the Hartree-Fock estimate of the effective mass translate into large corrections to the quasiparticle lifetime.

\subsection{Shear viscosity and thermal conductivity}

The shear viscosity coefficient of the fermion hard-sphere system of degeneracy $\nu$=2, $\eta$,  has been obtained
from Eqs. ~\eqref{eta_AK}, ~\eqref{lambda:eta} and ~\eqref{eta_sb} with the effective mass
and the in-medium scattering probability computed using  the CBF interaction.

\begin{figure}[htbp]
\begin{center}
\includegraphics[width= 0.45\textwidth]{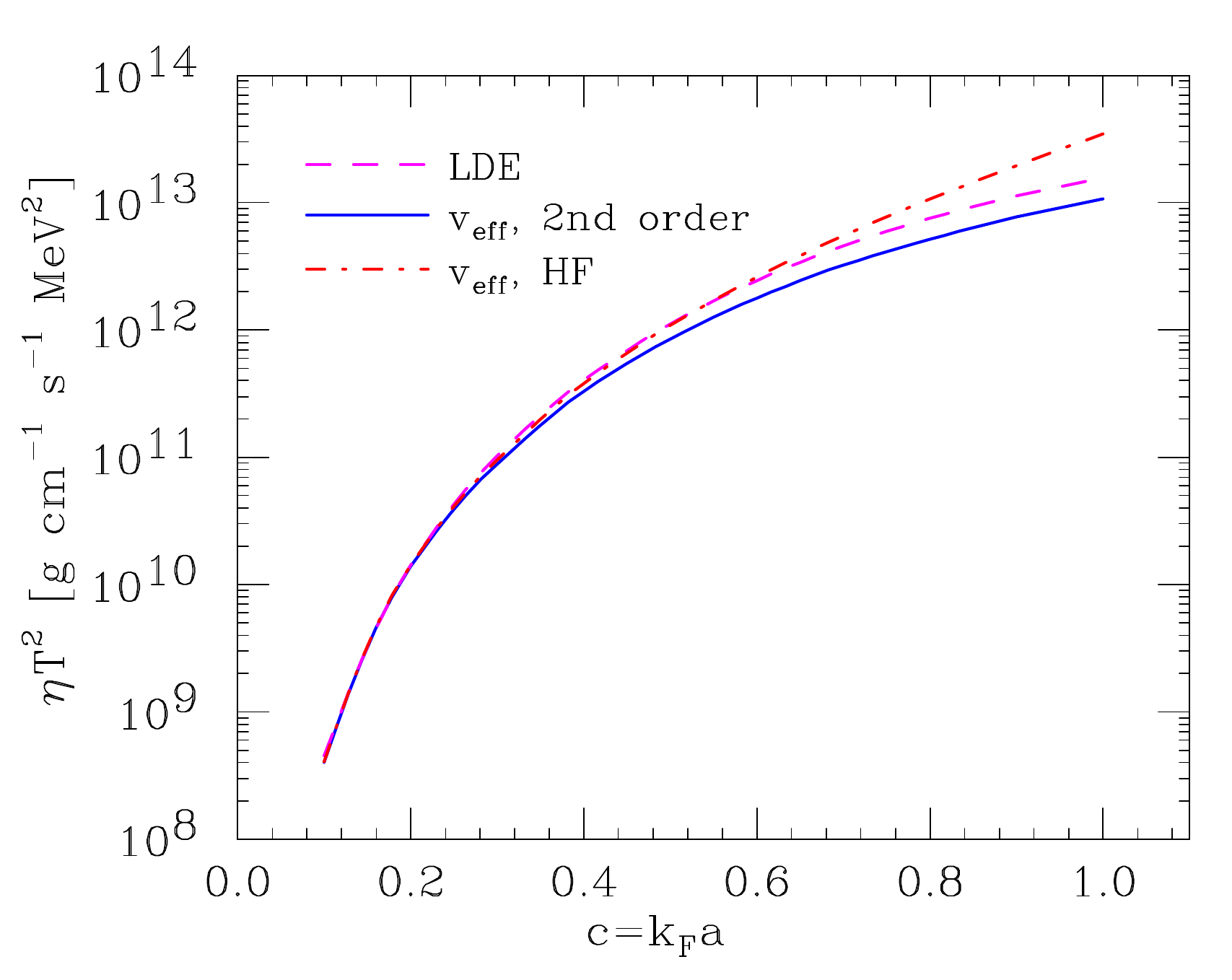}
\caption{(color online) $c$-dependence of the temperature-independent quantity $\eta T^2$, where
$\eta$ is the shear viscosity coefficient of the  fermion hard-sphere system of degeneracy  $\nu = 2$. The dot-dash and solid lines represent the results of calculations carried out using the first and second order approximations for the effective mass, respectively, while the dashed line shows the results of the low-density expansion of Ref. \cite{JLTP2}, including terms of order up to $c$.}
\label{eta1}
\end{center}
\end{figure}

Figure~\ref{eta1} shows the $c$-dependence of the
$T$-independent quantity   $\eta T^2$. The most relevant feature of the results displayed in the figure is, again,  the sizable effect
of second order contributions to the effective mass. As shown in Fig \ref{fig:mstar_nu2}, these corrections lead to sharp increase
of $m^\star$, which in turn implies  a decrease of the shear viscosity coefficient $\eta$.

\begin{figure}[htbp]
\begin{center}
\includegraphics[width= 0.45\textwidth]{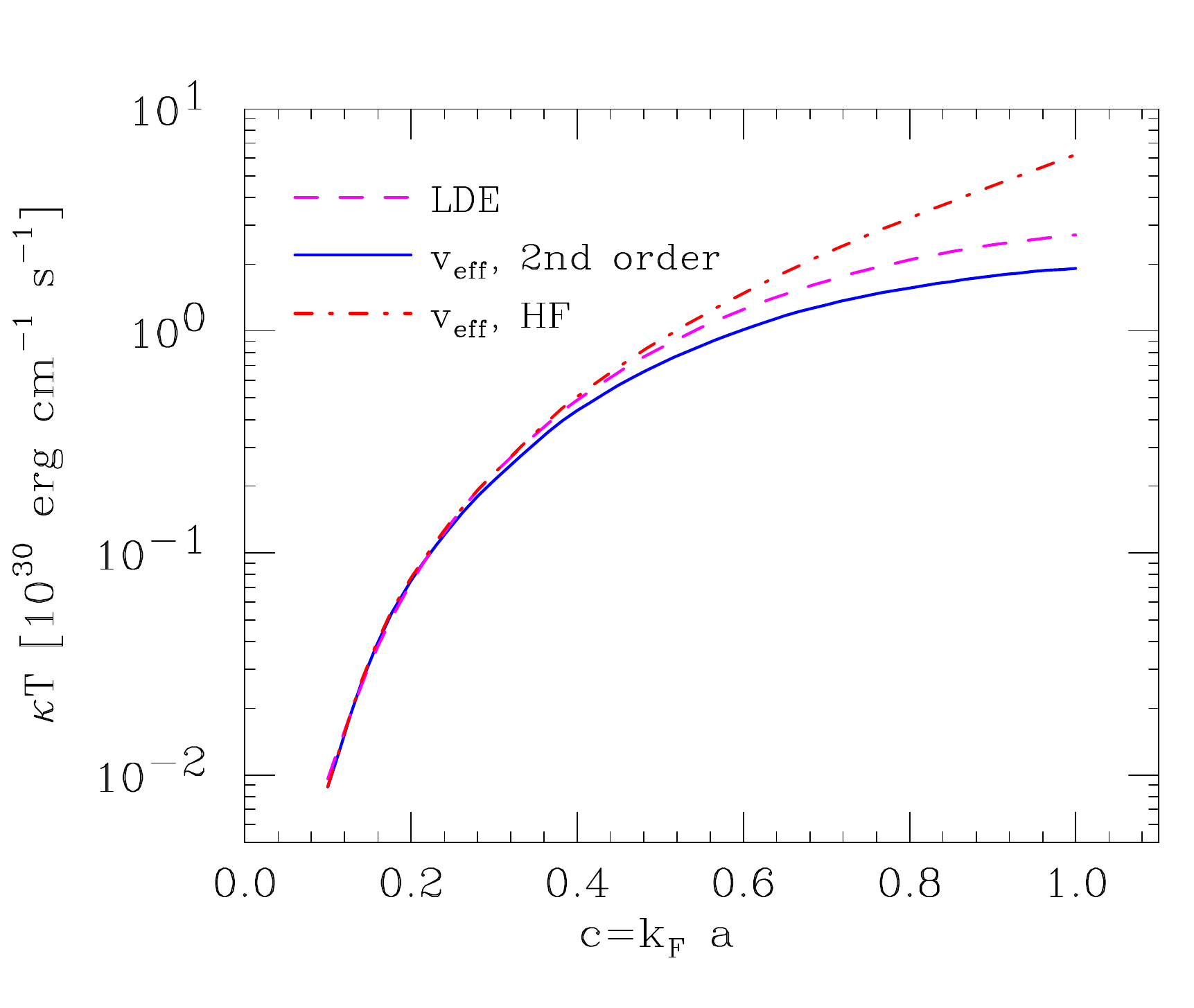}
\caption{(color online) $c$-dependence of the temperature-independent quantity $\kappa T$, where
$\kappa$ is the thermal conductivity of the fermion hard-sphere system of degeneracy  $\nu = 2$. The dot-dash and solid lines represent the results of calculations carried out using the first and second order approximations for the effective mass, respectively, while the dashed line shows the results of the low-density expansion of Ref. \cite{JLTP2}, including terms of order up to $c$.}
\label{cappa1}
\end{center}
\end{figure}

The $T$-independent quantity $\kappa T$,  where $\kappa$ is the
thermal conductivity defined by Eqs. ~\eqref{cappa_AK}, ~\eqref{lambda:eta} and ~\eqref{cappa_sb}, is shown in Fig. \ref{cappa1} as a
function of the dimensionless parameter $c$. Overall, the pattern is close to the one observed in Fig. \ref{eta1}.

\section{Summary and conclusions}
\label{conclusion}

We have extended the analysis of the hard-sphere fermion fluid performed in I, to study the shear viscosity and thermal conductivity
coefficients of the system with degeneracy $\nu$=2, that can be regarded as a model of pure neutron matter. 

\begin{figure}[htbp]
\begin{center}
\includegraphics[width= 0.35\textwidth]{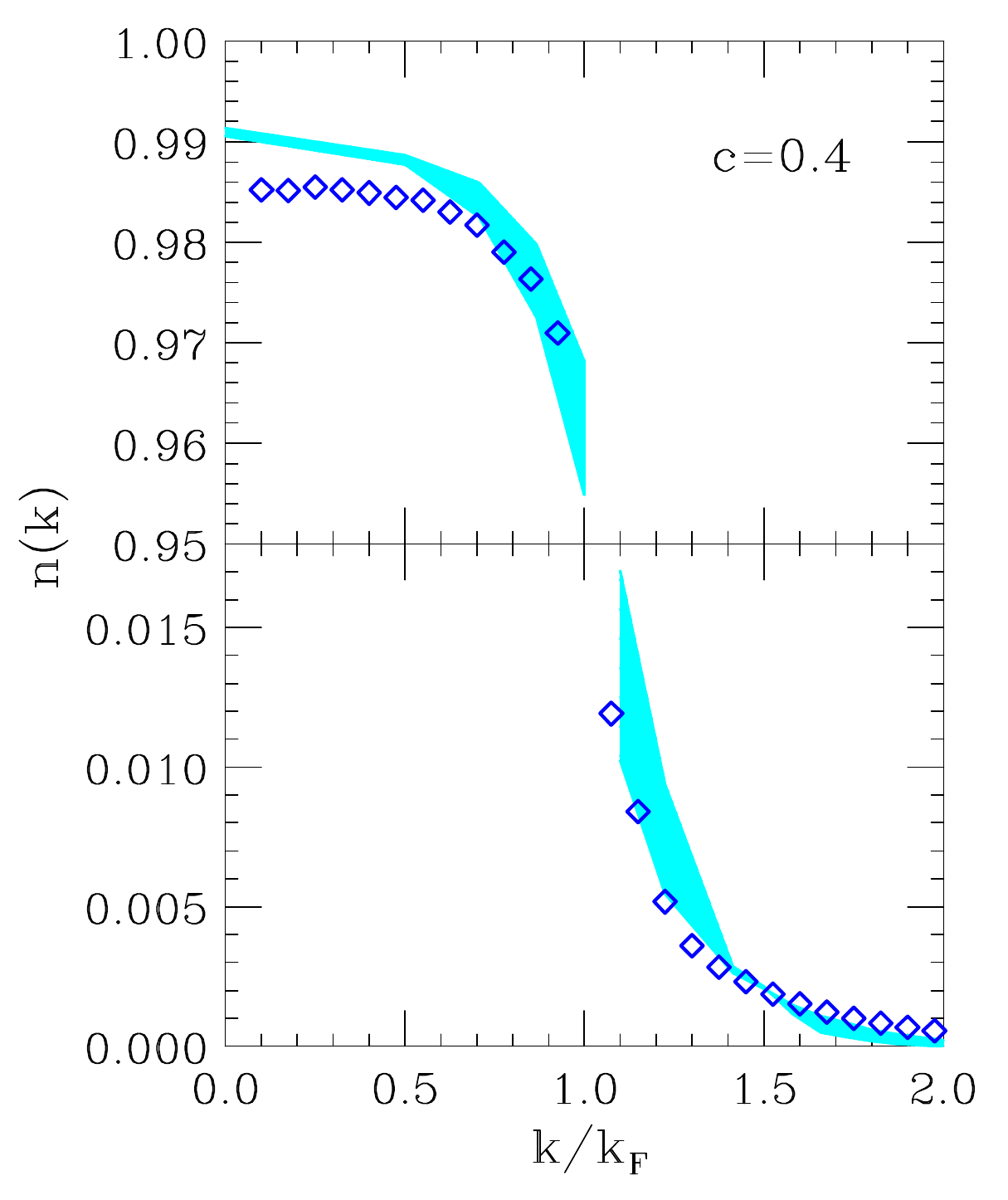}
\caption{(color online) Comparison between the momentum of the fermion hard-sphere system of degeneracy $\nu$=2 at $c=0.4$ (diamonds) and those
reported in Ref.~\cite{nk_MC}, corresponding to the density range $0.08  \leq  \rho \leq 0.24$ fm$^{-3}$  (shaded area). }
\label{nk}
\end{center}
\end{figure}

In order to establish a connection between our results and those corresponding to neutron matter, in Fig.~\ref{nk} we compare 
the momentum distribution of the fermion hard-sphere system at density corresponding to $c=0.4$\textemdash computed with the CBF effective interaction following the procedure described in I\textemdash to those reported in Ref.\cite{nk_MC}, 
obtained using a quantum Monte Carlo technique. The shaded region illustrates the variation of the momentum distribution 
of Ref.~\cite{nk_MC} in the density range  0.08$\leq \rho \leq$0.24 fm$^{-3}$. The corresponding values of the renormalisation constant 
are $Z$=0.9566 for the hard-sphere system and $0.9579 \leq Z \leq 0.9378$ for neutron matter. The appreciably higher values of $n(0)$
obtained from the Monte Carlo approach reflect the softness  of the chiral neutron-neutron potential employed by the authors 
of Ref.~\cite{nk_MC}.  The results of Fig.~\ref{nk} suggest that neutrons in pure neutron matter behave similarly to hard spheres of radius
$\lsim$ 0.3 fm. The same analysis for isoscalar nucleons, performed in I, leads to a radius of $\sim$0.4 fm .

The quantities needed for the calculations of the transport coefficients within the formalism developed by Abrikosov and Khalatnikov, i.e. 
the effective mass and in-medium scattering probability of particles on the Fermi surface, have been obtained using the 
approach based on the CBF effective interaction, originally derived in Refs.~\cite{shannon,BV}. This scheme alllows one to combine the 
flexibility of perturbation theory in the basis of eigenstates of the non interacting system with a realistic description of dynamical 
correlations in coordinate space.

Compared to the case of degeneracy $\nu$=4, discussed in I,  the derivation of the effective interaction for $\nu$=2 involves
additional difficulties arising from the presence of longer range statistical correlations. In the low-density region, corresponding 
to $c = k_F a < 0.5$, this problem manifests itself  in the lack of a clear minimum of the  FHNC energy as a function of the 
variational parameter describing the correlation range. 

To overcome this problem and test the validity of the variational FHNC approach employed in I, the reference value of the ground-state energy, $E_0$, 
needed to determine the CBF effective interaction from Eq.~~\eqref{newdef} has been calculated using both the VMC and DMC techniques. This analysis has 
confirmed that, in the density region in which they are available, the FHNC results provide an accurate upper bound to the ground-state energy of the 
fermion hard-sphere system for both $\nu$=2 and $\nu$=4. 

The in-medium scattering cross section, computed in Born approximation,  exhibits a significant density dependence and appears to smoothly approach 
the low-energy limit predicted by the partial wave expansion of the cross section in vacuum.

The real part of the proper self-energy computed at second order in the effective interaction has been employed to obtain the quasiparticle 
spectrum, the first derivative of which is simply related to the effective mass. As in the case $\nu$=4, the energy-dependent second order 
correction  significantly affects both the magnitude and the density dependence of the effective mass evaluated at $|{\bf k}| = k_F$.

The large second order effects on $m^\star(k_F)$ strongly impact on both the quasiparticle lifetime and the transport coefficients, leading to a 
suppression that monotonically increaeses with density. At density corresponding to $c$=$k_F a$=1, the ratio between the shear viscosity 
and thermal conductivity coefficients evaluated using the first and second order expressions of the effective mass turns out to be as large as 
$\sim3$, for $\eta$, and $\sim$5 for $\kappa$.  Such a big difference is likely to play an important role in astrophysical applications, and needs to be carefully investigated extending 
our analysis to nuclear matter. 

Owing to the complexity of the nucleon-nucleon interaction and to the presence three-nucleon forces,  which are known to be important 
at nuclear and supranuclear densities, the treatment of nuclear matter within the CBF effective interaction approach is somewhat more demanding 
from the computational point of view. However, it does not involve additional conceptual problems,  and allows for a consistent treatment 
of the variety of properties\textemdash ranging from the transport coefficients to the neutrino emissivity and mean free path, to the 
superconducting and superfluid gaps\textemdash that concur to determine many astrophysical processes. 

\acknowledgements
This research is supported by the U.S. Department of Energy, Office of Science, Office of
Nuclear Physics, under contract DE-AC02-06CH11357 (AL).  MICINN (Spain), under grant FI-2014-54672-P  and Generalitat de Catalunya, under grant 
2014SGR-401 (AP), and  INFN (Italy) under grant MANYBODY (AM and OB).  AM gratefully acknowledges the hospitality of the 
Departament d'Estructura i Constituents de la Mat\`eria  of the University of Barcelona, and support from ``NewCompStar'', 
COST Action MP1304.


\end{document}